\documentclass[prd,showpacs,floatfix,nofootinbib,twocolumn,amsmath,amssymb]{revtex4}
\usepackage{graphicx,graphics,color,dcolumn,booktabs,bm}
\usepackage{longtable,lscape}
\usepackage{txfonts}
\usepackage{overpic}
\usepackage{amssymb}
\usepackage{indentfirst}
\usepackage{epsfig}
\usepackage{feynmf}   
\usepackage{epstopdf}   
\usepackage{slashed}  
\usepackage{cases}
\usepackage{color}
\usepackage{multirow}

\usepackage[colorlinks, citecolor=blue,anchorcolor=red,menucolor=red, linkcolor=red,filecolor=red,runcolor=red,urlcolor=blue,frenchlinks=red]{hyperref}
\begin{document}

\title{Revealing the inner structure of the newly observed $D_2^*(3000)$}

\author{Jun-Zhang Wang$^{1,2}$}
\author{Dian-Yong Chen$^{3}$}\email{chendy@impcas.ac.cn}
\author{Qin-Tao Song$^{4,5}$}
\author{Xiang Liu$^{1,2}$\footnote{Corresponding author}}\email{xiangliu@lzu.edu.cn}
\author{Takayuki Matsuki$^{6,7}$}\email{matsuki@tokyo-kasei.ac.jp}
\affiliation{
$^1$Research Center for Hadron
and CSR Physics, Lanzhou University $\&$ Institute of Modern Physics
of CAS,
Lanzhou 730000, China\\
$^2$School of Physical Science and Technology, Lanzhou University,
Lanzhou 730000, China\\
$^3$Department of Physics, Southeast University, Nanjing 210094, People's Republic of China\\
$^4$ KEK Theory Center, Institute of Particle and Nuclear Studies, High Energy Accelerator Research Organization (KEK), 1-1, Ooho, Tsukuba, Ibaraki, 305-0801, Japan\\
$^5$Department of Particle and Nuclear Studies,
Graduate University for Advanced Studies (SOKENDAI),
1-1, Ooho, Tsukuba, Ibaraki, 305-0801, Japan\\
$^6$Tokyo Kasei University, 1-18-1 Kaga, Itabashi, Tokyo 173-8602, Japan\\
$^7$Theoretical Research Division, Nishina Center, RIKEN, Saitama 351-0198, Japan
}

\begin{abstract}
Stimulated by the $D_2^\ast(3000)$ recently observed by LHCb, we study the decay behaviors of the $3P$ and $2F$ charmed mesons. Comparing the masses and decay properties of the $3^3P_2$ and $2^3F_2$ with the observed $D_2^\ast(3000)$, we conclude that the most probable assignment is the $3^3P_2$, while the assignment of the $2^3F_2$ may not be fully excluded. The results of the unobserved $3P$ and $2F$ charmed mesons in this work may provide some fundamental information when further searching for these charmed mesons in the experiments by LHCb and the forthcoming Belle II.
\end{abstract}

 \maketitle

\section{introduction}\label{sec1}
Very recently, the LHCb Collaboration reported their amplitude analysis of $B^- \to D^+ \pi^- \pi^-$ decays in a data sample corresponding to $3.0\ \rm{fb}^{-1}$ of $pp$ collision data \cite{Aaij:2016fma}. The angular momenta of the resonances at the high $D^+ \pi^-$ invariant mass are analyzed, in which the contributions from the $D_2^\ast(2460)$, $D_1^\ast(2680)$ and $D_3^\ast(2760)$ are observed. Besides these three resonances, a higher one $D_2^\ast(3000)$ is also reported with the mass and width to be
\begin{eqnarray}
m      &=&3214 \pm 29 \pm 33 \pm 36 \ \rm{MeV},\\
\Gamma &=& 186 \pm 38 \pm 34 \pm 63 \ \rm{MeV},
\end{eqnarray}
respectively \cite{Aaij:2016fma}. The $J^P$ quantum number of the $D_2^\ast (3000)$ is measured to be $2^+$ \cite{Aaij:2016fma}.

Before the observation of the $D_2^*(3000)$, a great achievement was made in the past decades. A number of the charmed mesons were observed experimentally \cite{delAmoSanchez:2010vq, Aaij:2013sza, Aubert:2009wg, Link:2003bd, Abe:2003zm, Aubert:2006zb, Anjos:1988uf, Albrecht:1985as}. These observations not only made the charmed meson spectrum abundant, but also gave theorists motivation to reveal the nature of the newly observed states from the mass spectrum and decays \cite{ Godfrey:1985xj, Goity:1998jr, Di Pierro:2001uu, Bardeen:2003kt, Matsuki:1997da, Matsuki:2007zza, Matsuki:2006rz, Matsuki:2011xp, Ebert:1997nk}, which could undoubtedly deepen our understanding of the highly excited charmed mesons.

Thirty years ago, Godfrey and Isgure (GI) performed a systematical study of the meson spectra in a semirelativistic quark model \cite{Godfrey:1985xj}, in which a quenched confinement potential was employed to describe the quark-antiquark interactions. The low-lying charmed mesons were well reproduced by the so-called GI model \cite{Godfrey:1985xj}. However, there exist some difficulties in understanding the recently observed higher excitations of the charmed mesons, which are due to a typical quenched model of the GI model. As proposed in Refs. \cite{Song:2015nia, Song:2015fha}, we include the unquenched peculiarity in the semirelativistic quark model by replacing the linear confining potential with a screened one, which is consistent with the unquenched lattice QCD calculation and some holographic model \cite{Bali:2005fu, Armoni:2008jy, Bigazzi:2008gd, Namekawa:2011wt}. In this modified GI model \cite{Song:2015nia, Song:2015fha}, the charmed and charmed strange meson spectra and their decays could be better described than the GI model.

\begin{figure}[t]
\scalebox{0.70}{\includegraphics{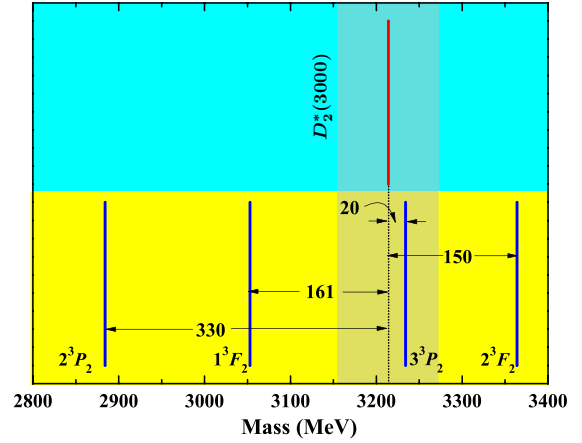}}
\caption{A comparison of the masses of the $2^3P_2$, $3^3P_2$, $1^3F_2$ and $2^3F_2$ charmed mesons of the modified GI model  \cite{Song:2015fha} and the experimental measurement of the $D_2^\ast(3000)$ \cite{Aaij:2016fma}. The light gray band indicates the error of the $D_2^\ast(3000)$ mass \cite{Aaij:2016fma}. \label{Fig:spectrum}}
\end{figure}

As for the newly observed $D_2^\ast(3000)$, it could be a $P$-wave or $F$-wave charmed meson. With the same parameters as in \cite{Song:2015fha}, the masses of the $2^3P_2$, $3^3P_2$, $1^3F_2$, and $2^3F_2$ charmed mesons are calculated, and are shown in Fig. \ref{Fig:spectrum}. As a comparison, the newly observed $D_2^\ast(3000)$ is also presented and the mass differences of the theoretical calculation and the central value of the observed one are also marked. The mass of the $D_2^\ast(3000)$ is quite consistent with the predicted mass of a $3^3P_2$ state, which is $3234$ MeV. As for the mass of a $2^3P_2$ state, the theoretical estimate is 2884 MeV, which is about 330 MeV below the central value of the $D_2^\ast(3000)$ mass. In addition, the total width of the $2^3P_2$ state was estimated to be 68.89 MeV when taking its mass as 2884 MeV in Ref. \cite{Song:2015fha}, which is much smaller than the $D_2^\ast(3000)$. As for the $F$-wave charmed meson, the estimated masses of the $1^3F_2$ and $2^3F_2$ states are 3053 and 3364 MeV, respectively, and are about 161 MeV below and 150 MeV above the $D_2^\ast(3000)$. The total width of the $1^3F_2$ was estimated to be 222.02 MeV when taking its mass as 3053 MeV in Ref. \cite{Song:2015fha}, which does not conflict with the $D_2^\ast(3000)$. Although the observed $D_2^\ast(3000)$ could be a good candidate of the $3^3P_2$ from the spectral point of view, we cannot fully exclude the other assignments of the $D_2^\ast(3000)$ such as a $2^3F_2$ state considering the uncertainty of the theoretical predictions.

To further check possibilities of different assignments of the $D_2^\ast(3000)$, we study its decay behaviors in different cases. In the present work, we adopt the quark pair creation (QPC) model to estimate the Okubo-Zweig-Iizuka (OZI) allowed decays of the $D_2^\ast(3000)$, which was first proposed by Micu \cite{Micu:1968mk}, and further developed by the Orsay group \cite{ LeYaouanc:1972vsx, LeYaouanc:1973ldf, LeYaouanc:1974cvx, LeYaouanc:1977fsz, LeYaouanc:1977gm}. The decay behaviors of the $2P$ and $1F$ charmed mesons have been investigated in Ref. \cite{Song:2015fha}. In the present work, we further study the decays of the $3P$ and $2F$ states to reveal the inner structure of the newly observed $D_2^\ast(3000)$. The decay behaviors of the $D_2^\ast(3000)$ could in turn check our understanding of the charmed meson spectrum proposed in Ref. \cite{Song:2015fha}.

This paper is organized as follows. In Sec. \ref{sec2} we will briefly review the QPC model and present our calculation of the two-body OZI-allowed decays of the $3P$ and $2F$ charmed mesons. Section \ref{sec3} is dedicated to a discussion and summary.

\section{QPC model and decays of the $3P$ and $2F$ charmed mesons \label{sec2}}
The QPC model is a phenomenological method to estimate the OZI allowed strong decays. In the QPC model, the quark-antiquark pair is created from the vacuum with $J^{PC}=0^{++}$. Thus the QPC model is also named the $^3P_0$ model. In the following, we present a brief review of the QPC model. Taking the OZI allowed strong decay process $A\to BC$ as an example, the corresponding $S-$ matrix is 
\begin{eqnarray}
\langle BC \left|S \right| A\rangle = I-i (2\pi) \delta(E_f -E_i)  \langle BC \left|T \right| A\rangle.
\end{eqnarray}
In the nonrelativistic limit, the transition operator $T$ is defined as
\begin{eqnarray}
T& = &-3\gamma \sum_{m,i}\langle 1m;1-m|00\rangle\int d \mathbf{p}_3d\mathbf{p}_4 \delta ^3 (\mathbf{p}_3+\mathbf{p}_4) \nonumber \\
&& \times \mathcal{Y}_{1m}\left( \frac{\mathbf{p}_3-\mathbf{p}_4}{2} \right) \chi _{1,-m}^{34} \phi _{0}^{34}
\omega_{0}^{34} b_{3i}^{\dag} (\mathbf{p}_3) d_{4i}^{\dag}(\mathbf{p}_4).
\end{eqnarray}
This transition operator is introduced to describe the quark-antiquark (denoted by indices 3 and 4, respectively) created from the vacuum. The phenomenological creation strength $\gamma$ for $q\bar{q}$ is taken as $\gamma=8.7$ \cite{Song:2015fha}, while the strength for $s\bar{s}$ satisfies $\gamma_s=\gamma/\sqrt{3}$. $\mathcal{Y}_{\ell m}(\mathbf{p})=|\mathbf{p}|^\ell Y_{\ell m}({\hat{\mathbf{p}}}) $ is the $\ell$th solid harmonic polynomial. $\chi$, $\phi$, and $\omega$ are the general description of the spin, flavor, and color wave functions of the quark-antiquark pair, respectively.

With the above transition operator $T$, the transition matrix of a process $A\to BC$ is given by 
\begin{eqnarray}
\langle BC | T| A\rangle = \delta^3(\mathbf{p}_B+\mathbf{p}_C) \mathcal{M}^{M_{J_A}M_{J_B}M_{J_c}}
\end{eqnarray}
where $\mathbf{P}_B$ and $\mathbf{P}_C$ are the momenta of the mesons $B$ and $C$ in the $A$ rest frame, respectively. $|A\rangle$, $|B\rangle$ and $|C\rangle$ denote mock states corresponding to mesons $A$, $B$ and $C$, respectively \cite{Hayne:1981zy}.  The mock state $|A\rangle$ is defined as s
\begin{eqnarray}
&&|A(n^{2S+1}L_{JM_J})(\mathbf{p}_A)\rangle\nonumber\\
&&=\sqrt{2E}\sum\limits_{ {M_{S}},{M_{L}}}\langle LM_L;SM_{S}|JM_{J}\rangle\chi^{A}_{S,M_s} \nonumber\\
&&\quad\times \phi^A \omega^A\int d\mathbf{p}_1 d\mathbf{p}_2 \delta^3(
\mathbf{p}_A-\mathbf{p}_1-\mathbf{p}_2) \nonumber\\ &&\quad\times\Psi^A_{nLM_L}(\mathbf{p}_1,\mathbf{p}_2)|q_1(\mathbf{p}_1) \bar{q}_2(\mathbf{p}_2)\rangle,
\end{eqnarray}
where $\chi^A$, $\phi^A$, $\omega^A$ and $\Psi^A$ are the spin, flavor, color and spatial wave functions of the meson $A$. In the present work, we use the spatial wave functions of the charmed and charmed strange mesons estimated in Refs. \cite{Song:2015fha, Song:2015nia}, while the one for a light meson is calculated from the GI model \cite{Godfrey:1985xj}.

The partial wave amplitudes are related to the helicity amplitudes by \cite{Jacob:1959at}
\begin{eqnarray}
\mathcal{M}^{JL}(\mathbf{P})&=&\frac{\sqrt{2L+1}}{2J_A+1}\sum_{M_{J_B}M_{J_C}}\langle L0;JM_{J_A}|J_AM_{J_A}\rangle \nonumber \\
&&\times \langle J_BM_{J_B};J_CM_{J_C}|{J_A}M_{J_A}\rangle \mathcal{M}^{M_{J_{A}}M_{J_B}M_{J_C}},
\end{eqnarray}
and the partial width of the $A\to BC$ is
\begin{eqnarray}
\Gamma_{A\to BC} &=& \pi ^2\frac{|\mathbf{P}_B|}{m_A^2}\sum_{J,L}|\mathcal{M}^{JL}(\mathbf{P})|^2,
\end{eqnarray}
where $m_{A}$ is the mass of the initial state $A$.

\begin{table}[htbp]
\caption{The masses of the charmed and charmed-strange mesons involved in the present calculation. For the mass of the $D^{(\ast)}$ , we use the average of $m_{D^{(\ast)0}}$ and $m_{D^{(\ast)+}}$. }
\label{mass}
\begin{center}
\begin{tabular}{cccccc}
\toprule[1pt]
\midrule[1pt]
 State & Mass (MeV) & State & Mass (MeV) \\
\midrule[1pt]
$D$               &$1867$ \cite{Agashe:2014kda}    &
$D(1D_2)$         &$2761$ \cite{Song:2015fha} \\
$D^*$             &$2009$ \cite{Agashe:2014kda}    &
$D(1D^{'}_2)$     &$2791$ \cite{Song:2015fha} \\
$D_0^*(2400)$     &$2318$ \cite{Agashe:2014kda}
& $D_s$           &$1968$ \cite{Agashe:2014kda}\\
$D_1(2420)$       &$2421$ \cite{Agashe:2014kda}  &
$D_s^*$           &$2112$ \cite{Agashe:2014kda}\\
$D_1(2430)$       &$2427$ \cite{Agashe:2014kda}   &
$D_{s0}^*(2317)$  &$2318$ \cite{Agashe:2014kda}\\
$D_2^*(2460)$     &$2463$ \cite{Agashe:2014kda}      &
$D_{s1}(2460)$    &$2460$ \cite{Agashe:2014kda}\\
$D(2550)$         &$2539$ \cite{Agashe:2014kda}&
$D_{s1}(2536)$    &$2535$ \cite{Agashe:2014kda}\\
$D^*(2600)$       &$2612$ \cite{Agashe:2014kda}&
$D_{s2}^*(2573)$  &$2572$ \cite{Agashe:2014kda}\\
$D(1^3D_1)$       &$2762$ \cite{Song:2015fha}  &
$D_s(2^1S_0)$     &$2644$ \cite{Song:2015nia}\\
$D(1^3D_3)$       &$2779$ \cite{Song:2015fha} &
$D_{s1}^*(2700)$  &$2709$ \cite{Agashe:2014kda}\\
\midrule[1pt]
\bottomrule[1pt]
\end{tabular}
\end{center}
\end{table}

The masses of the involved charmed and charmed strange mesons are listed in Table \ref{mass}. For the low-lying observed mesons, we adopt the central values of the PDG average, while for the unobserved higher excited charmed and charmed strange mesons, we use the theoretical mass estimate in the modified GI model as an input \cite{Song:2015fha, Song:2015nia}. As for light mesons, we use the same masses as those summarized in Table III of Ref. \cite{Yu:2011ta}\footnote{In our treatment of charmed and charmed strange mesons, we consider the screening effect since we are discussing higher excited states in charmed and charmed strange meson families, where coupled-channel effects become important. Different from the case of charmed and charmed strange mesons, we use the spatial wave function with the unscreening effect for the light mesons involved in the discussed decay modes since coupled-channel effects on the light mesons that are ground states are not obvious. Thus, we adopt this approximation to deal with light mesons.}. Besides the masses of the involved mesons, in the present calculations, the masses of the quarks are taken as 1.628, 0.419 and 0.22 MeV for the charmed, strange, up/down quarks, respectively.

\begin{table*}[htbp]
\caption{ The partial and total widths of $3^3P_0$, $3^3P_2$, $2^3F_2$ and $2^3F_4$ charmed mesons in units of MeV.}
\label{Tab:decay}
\begin{center}
\begin{tabular}{cccccccccc}
\toprule[1pt]
\toprule[1pt]
Channels      &   $3^3P_0$&    $3^3P_2$&      $2^3F_2$ &    $2^3F_4$ &  Channels   &   $3^3P_0$&    $3^3P_2$&   $2^3F_2$ &  $2^3F_4$  \\
              &    3219 MeV &   3234 MeV &     3364 MeV&    3345 MeV &             &   3219 MeV &   3234 MeV &  3364 MeV &  3345 MeV   \\
\midrule[1pt]
\\
  $D \pi$     &  22.6        &4.28     &14.2   & 2.81                   &$D_1(2420) \omega$     &2.02    &0.592          &3.81   &0.566        \\
  $D \eta$     & 2.35        &0.994     &2.58   &0.216                    &$D_1(2430) \pi$       &2.17    &2.05           &6.24   & 1.26          \\
  $D \rho$     & -           &9.94     &4.44   & 0.688                    & $D_1(2430) \eta$     &0.339       &0.0466       &0.975  & 0.405           \\
  $D \omega$     & -         &3.22      & 1.44  & 0.241                   & $D_1(2430) \sigma$    &-         &0.0248        &0.0654 & 0.000664        \\
  $D \eta^{'}$     &0.0155   &0.318      &0.757 &$1.15\times10^{-5}$       &$D_1(2430) \rho$       &0.689      &3.75          &2.57   &2.04       \\
  $D h_1(1170)$     &6.69     & 1.12    &2.98   &1.84                   &$D_1(2430) \omega$     &0.0833       &0.895         &0.830  &0.612    \\
  $D b_1(1235)$     &33.9     & 2.84     &3.86 & 4.58                 &$D_2^*(2460) \pi$          &-    &5.32                 &5.25     & 2.29         \\
  $D a_1(1260)$     &0.0220   & 2.6      &0.642 & 2.39                &$D_2^*(2460) \eta$       &-       &0.422               &0.886    & 0.352        \\
  $D f_2(1270)$     & -       & 1.47     &1.69  & 2.44              &$D_2^*(2460) \sigma$     &3.35        &1.76         & 0.255       & 1.06              \\
  $D f_1(1285)$     & 0.0757   & 0.623    &0.0238 & 0.370         &$D_2^*(2460) \rho$      &-        &-              &4.97            & 21.4         \\
  $D \eta(1295)$     &7.08     & 0.752    &3.30   & 0.325        & $D_2^*(2460) \omega$   &-         &-             &1.58           & 7.45     \\
  $D \pi(1300)$     & 39.4     & 3.35      &16.3  & 1.49         &$D (2550) \pi$        &20.7     &0.637          &2.65          & 4.12        \\
  $D a_2(1320)$     & -       & 3.49      &4.46  & 4.72        &$D (2550) \eta$       &0.0246     &0.503         &1.92         & 0.243  \\
  $D f_1(1420)$     & -       & -         &0.493 & 0.00684        &$D (2550) \rho$       &-         &-             &1.88       & 0.00202           \\
  $D^* \pi$        & -        &10.4          &8.76 &1.47       & $D (2550) \omega$    &-         &-             & 0.461        &0.000212   \\
  $D^* \eta$       &-        &1.81          &1.36 &0.0414        & $D^* (2600) \pi$     &-         &3.70           &3.81       &4.78   \\
  $D^* \sigma$     &2.17    &0.427         &1.01 & 1.40        &$D^* (2600) \eta$     &-         &0.360          &1.36        & 0.126   \\
  $D^* \rho$     &6.12        &8.13        &25.4 &18.9         &$D^* (2600) \sigma$   &0.116      &0.170         &0.648         &0.00522       \\
  $D^* \omega$     &1.96    &2.65          &8.44 &6.12        &$D(1^3D_1) \pi$       &  -       &0.717         &0.653          &0.115       \\
  $D^* \eta^{'}$     & -     & 0.131       &0.138&0.0284        &$D(1^3D_1) \eta$      &  -       &-             &0.0100      &$7.19\times10^{-6}$         \\
  $D^* f_0(980)$     &0.342  &0.225        &1.95 &0.424        &$D(1^3D_1) \sigma$    &  -       &-            &0.00360      & -     \\
  $D^* a_0(980)$     &2.75    &1.75         &14.1 &3.01       &$D(1^3D_3) \pi$       &  -       &3.69          &6.49         & 3.02       \\
  $D^* h_1(1170)$     &0.365  &1.8         &1.97 & 1.54       &$D(1^3D_3) \eta$      &  -       &-             &0.0567       &0.00577             \\
  $D^* b_1(1235)$     & -     & -          & 9.59 &1.77       &$D(1D_2) \pi$         &29.9       &1.49          &38.4       &0.694        \\
  $D^* a_1(1260)$     & -    & -           & 11.4 &0.978       &$D(1D_2) \eta$        &  -       &-             &2.62      &$4.03\times10^{-5}$       \\
  $D^* f_2(1270)$     & -   & -            & 2.35 &15.8        &$D(1D_2) \sigma$      &  -       &-            &0.00479    & -      \\
  $D^* f_1(1285)$     & -     & -          & 1.52 &0.0344       & $D(1D_2^{'}) \pi$    &6.64       &3.16          &2.57     & 4.65        \\
  $D^* \eta(1295)$     & -    &-           & 0.574 &0.00248      &$D(1D_2^{'}) \eta$    &  -       &-           &0.195      & 0.000516     \\
  $D^* \pi(1300)$     & -     & -           & 2.46 &0.00741      &$D_s  K$              & 0.128    &0.455        & 0.911     & 0.00192      \\
  $D^* a_2(1320)$     & -     & -           & 3.82 &18.0             &$D_s  K^*$            & -        &0.0771       & 0.00440   &0.0792       \\
  $D_0^*(2400) \sigma$ &1.13   & 0.228       & 0.523 &0.0168         &$D_s^*  K$            &-       &0.576         & 0.352     & 0.0162  \\
  $D_0^*(2400) \rho$ &6.70      & 2.72        &3.94  &1.91              &$D_s^*  K^*$           &1.70      &1.90      & 0.782    & 0.115     \\
  $D_0^*(2400) \omega$ &2.28    & 0.953       &1.28  &0.617               &$D_{s0}^*(2317)  K^*$  &0.0478    &0.109        &0.166    &0.0492     \\
  $D_0^*(2400) f_0(980)$ & -   &-           &0.170 &$2.75\times10^{-5}$      &$D_{s1}(2460)  K$      &0.0923       &0.193   & 1.24    &0.0631     \\
  $D_0^*(2400) a_0(980)$ & -   &-           &1.08  & 0.000130              &$D_{s1}(2460)  K^*$    & -           &-         & 0.0858  &-        \\
  $D_1(2420) \pi$     &34.4    & 0.323        &31.1  &2.92                   &$D_{s1}(2536)  K$     &0.796    &0.0401       & 0.322   &0.187     \\
  $D_1(2420) \eta$     &0.550  &0.311       &2.84  &0.188                   &$D_{s2}^*(2573)  K$    & -        &0.0735       & 0.292   &0.124     \\
  $D_1(2420) \sigma$    &-    &0.479       &0.763 &0.00653                  &$D_{s}(2^1S_0)  K$    &1.82        &0.103        & 1.15    &0.0279  \\
  $D_1(2420) \rho$    &9.61     &2.20         &11.5  &1.89                 &$D_{s1}^*(2700)  K$  & -        &0.0161       & 0.583   &0.0112        \\\\
  \midrule[1pt]
  Total                 &251.1     &102.4    & 302.2  & 155.0        \\

\bottomrule[1pt]
\bottomrule[1pt]
\end{tabular}
\end{center}
\end{table*}

\subsection{Decay behaviors of the $3P$ charmed mesons}

To further test the possibility of the $D_2^\ast(3000)$ as the $3^3P_2$ charmed meson, we investigate the decay behaviors of the $3^3P_2$ charmed meson in the QPC model. The total and partial widths of the $3^3P_2$ state are listed in Table. \ref{Tab:decay}. The total width of the $3^3P_2$ charmed meson is estimated to be 102.4 MeV, which is consistent with the lower limit of the $D_2^\ast(3000)$ width, 105 MeV. In addition, the $3^3P_2$ charmed meson dominantly decays into $D^\ast \pi$, $D \rho$, $D^\ast \rho$, $D_2^\ast(2460) \pi$, and $D\pi$, the corresponding partial decay widths of which are 10.4, 9.94, 8.13, 5.32 and 4.28 MeV, respectively. The $3^3P_2 \to D \pi$ mode is the fifth dominant decay channel, which is also consistent with the experimental measurement of the $D_2^\ast(3000)$ in the $D\pi$ invariant mass spectrum. In addition, the ratios of the partial widths of the $D^\ast \pi$, $D \rho$, $D^\ast \rho$ and $D_2^\ast(2460) \pi$ modes and the one of the $D\pi$ channel are
\begin{eqnarray}
\frac{\Gamma[3^3P_2 \to D^\ast \pi]}{\Gamma[3^3P_2 \to D \pi]} &=& 2.42,\nonumber\\
\frac{\Gamma[3^3P_2 \to D \rho]}{\Gamma[3^3P_2 \to D \pi]} &=& 2.22,\nonumber\\
\frac{\Gamma[3^3P_2 \to D^\ast \rho]}{\Gamma[3^3P_2 \to D \pi]} &=& 1.90,\nonumber\\
\frac{\Gamma[3^3P_2 \to D_2^\ast(2460) \pi]}{\Gamma[3^3P_2 \to D \pi]} &=& 1.24,
\end{eqnarray}
respectively, which could be tested by future experiments.

There exist four $3P$ charmed mesons; the $D_2^\ast(3000)$ could be a candidate of the $3^3P_2$ charmed meson. The mass of the $3^3P_0$ charmed meson is estimated to be $3219$ MeV in the modified GI model \cite{Song:2015fha}. The total and partial widths of the $3^3P_0$ charmed meson are listed in Table \ref{Tab:decay}. The total width of the $3^3P_0$ charmed meson is estimated to be 251 MeV, which indicates that the $3^3P_0$ charmed meson is a broad state. The dominant decay modes of the $3^3P_0$ state are $D\pi(1300)$, $D_1(2420) \pi$, $Db_1(1235)$, $D(1D_2) \pi$, $D \pi$, and $D(2550) \pi$, the corresponding partial widths of which are 39.4, 34.4, 33.9, 29.9, 22.6, and 20.7 MeV, respectively. The branching ratio of the $3^3P_2 \to D\pi$ is about $9 \%$, which indicates that the $3^3P_0$ charmed meson can be detected by the $D\pi$ mode.

\begin{figure}[htbp]
\begin{center}
\scalebox{0.5}{\includegraphics{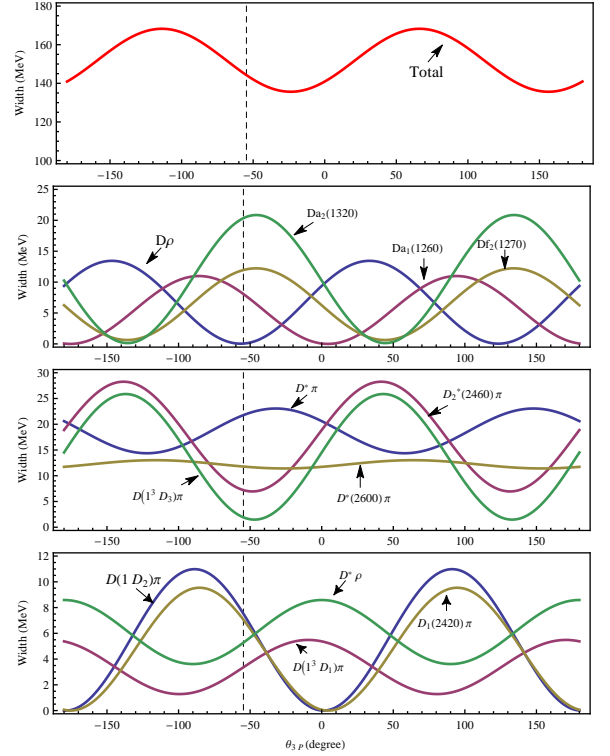}}
\caption{ Mixing angle $\theta_{3P}$ dependence of the total and partial widths of the $3P(1^+)$ charmed meson. The vertical dashed line indicates the mixing angle $\theta_{3P}$ in the heavy quark limit, which is $-54.7^{\circ}$. \label{Fig:3P}}
 \end{center}
\end{figure}

\begin{figure}[htbp]
\begin{center}
\scalebox{0.5}{\includegraphics{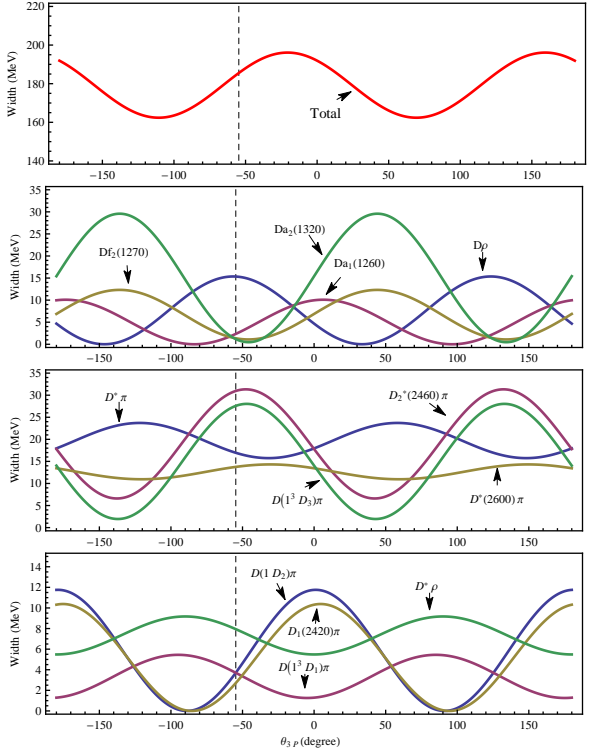}}
\caption{The same as Fig. \ref{Fig:3P} but for the $3P^\prime(1^+)$ charmed meson. \label{Fig:3Pprime} }
 \end{center}
\end{figure}

The two $1^+ $ charmed mesons are the mixtures of the $^3P_1$ and $^1P_1$ states and the mixing scheme for the $3P$ charmed meson is
\begin{eqnarray}
\left(
\begin{array}{c}
  |3P(1^+)\rangle \\
  |3P^\prime(1^+) \rangle
\end{array}\right)
=
\left(
\begin{array}{cc}
 \cos(\theta_{3P}) & \sin (\theta_{3P}) \\
 -\sin(\theta_{3P}) &\cos (\theta_{3P}) \\
\end{array}
\right)
\left(
\begin{array}{c}
  |3^1P_1 \rangle \\
  |3^3P_1 \rangle
\end{array}
\right),
\end{eqnarray}
where $\theta_{3P}$ is the mixing angle of the $|3^1P_1\rangle$ and $|3^3P_1\rangle$ states. In the modified GI model, the masses of the $|3^1P_1\rangle$ and $|3^3P_1\rangle$ states are estimated to be 3230 and 3215 MeV, respectively \cite{Song:2015fha}.

The $\theta_{3P}$ dependence of the total and partial widths of the $3P(1^+)$ charmed mesons is presented in Fig. \ref{Fig:3P}. In the heavy quark limit, the mixing angle of the $3P$ charmed mesons is $\theta_{3P}=  -\arcsin(\sqrt{2/3})=-54.7^\circ$ \cite{Godfrey:1986wj, Matsuki:2010zy, Barnes:2002mu}.
With this mixing angle, we find the total width of the $3P(1^+)$ charmed meson is 144 MeV. Our present calculations indicate that $Da_2(1320)$ and $D^\ast \pi$ are the dominant decay channels of the $3P(1+)$ charmed meson, the corresponding partial widths of which are 20.4 and 21.5 MeV, respectively. In addition, the decay modes $D f_2(1270)$ and $D^\ast(2600) \pi$ are also important since the corresponding partial widths are estimated to be about 10 MeV. The $\theta_{3P}$ dependence of the total and partial widths of the $3P^\prime(1^+)$ is presented in Fig. \ref{Fig:3Pprime}. With the mixing angle in the heavy quark limit, the total width of the $3P^\prime(1^+)$ is predicted to be 185 MeV, which is a bit larger than the one of the $3P(1^+)$ charmed meson. The dominant decay modes of the $3P^\prime(1^+)$ charmed meson are $D_2^\ast(2460) \pi$ and $D(1^3D_3) \pi$, the corresponding partial widths of which are 30.9 and 27.4 MeV, respectively. The partial widths of the $D^\ast \pi$, $D\rho$ and $D^\ast(2600) \pi$ are all about 15 MeV, which are also important decay modes of the $3P^\prime(1^+)$ charmed meson.

\begin{figure}[htbp]
\begin{center}
\begin{tabular}{c}
\scalebox{0.5}{\includegraphics{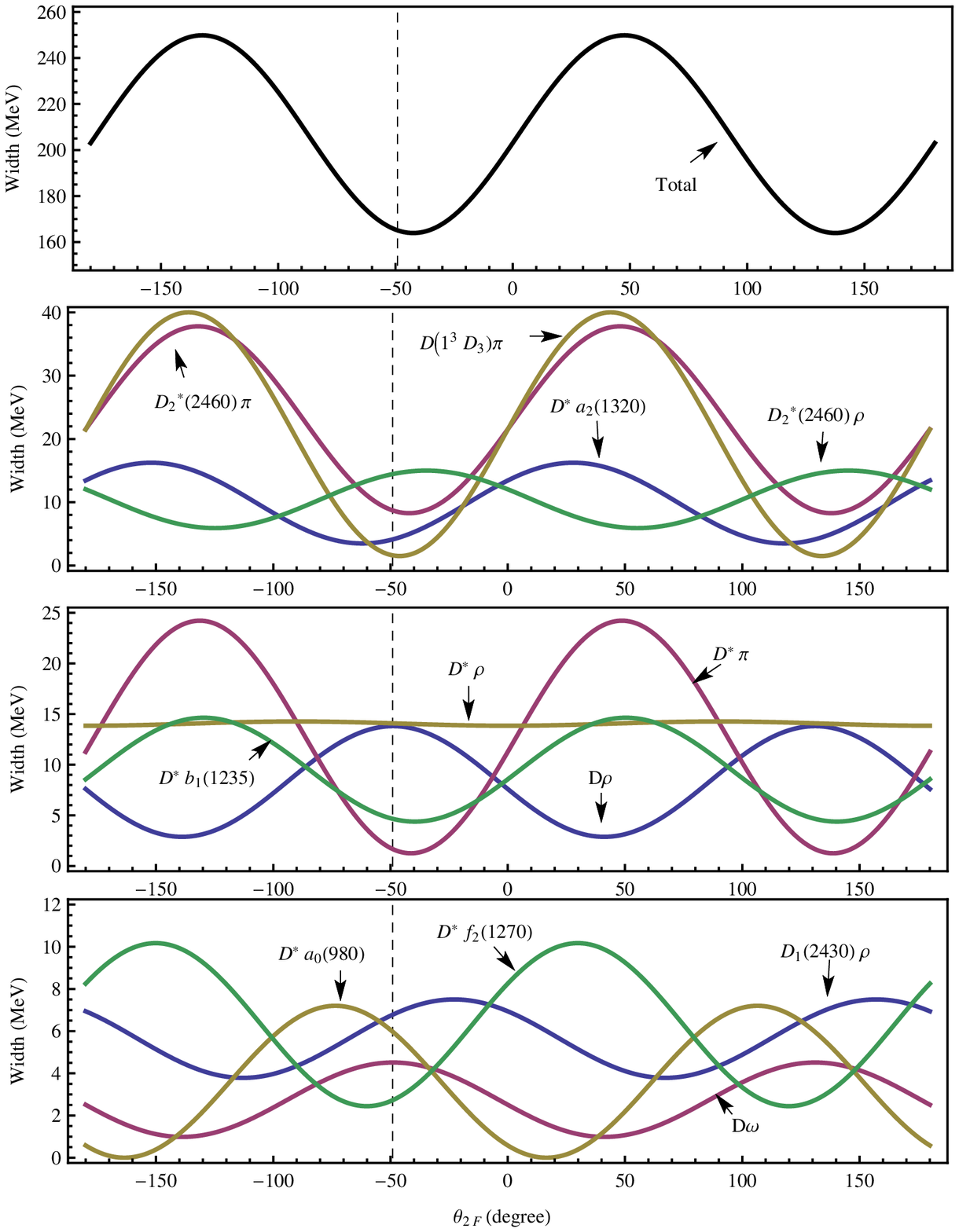}}
\end{tabular}
\caption{Mixing angle $\theta_{2F}$ dependence of the total and partial widths of the $|2F(3^+)\rangle$ charmed meson. The vertical dashed line indicates the mixing angle $\theta_{2F}$ in the heavy quark limit, which is $-49.1^{\circ}$. \label{Fig:2F} }
 \end{center}
\end{figure}

\begin{figure}[htbp]
\begin{center}
\begin{tabular}{c}
\scalebox{0.5}{\includegraphics{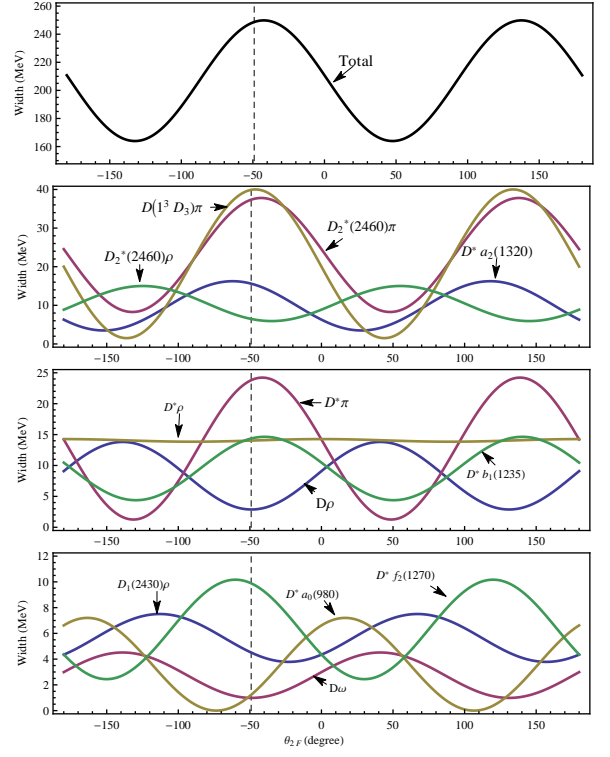}}
\end{tabular}
\caption{The same as Fig. \ref{Fig:2F} but for $|2F^\prime(3^+)\rangle$ charmed meson.\label{Fig:2Fprime} }.
 \end{center}
\end{figure}

\subsection{Decay behaviors of the $2F$ charmed mesons}

The $2F$ charmed mesons have not been observed, and the predicted mass of the $2^3F_2$ state is about 150 MeV above the central value of the mass of the newly observed $D_2^\ast(3000)$ \cite{Aaij:2016fma}. In the present work, we also consider the possibility of the $D_2^\ast(3000)$ as the $2^3F_2$ charmed meson. The total and partial widths of the $2^3F_2$ charmed meson are listed in Table \ref{Tab:decay}. The total width of the $2^3F_2$ charmed meson is predicted to be 302.2 MeV, which is about 35 MeV above the upper limit of the observed value of the $D_2^\ast(3000)$. The partial width of the $D\pi $ channel is 14.2 MeV, and the corresponding branching ratio is $4.7 \%$, which is almost the same as that of $3^3P_2 \to D\pi$. The $D(1D_2)\pi$, $D_1(2420) \pi$, $D^\ast \rho$, $D\pi(1300)$ and $D^\ast a_0(980)$ are also dominant decay channels of the $2^3F_2$ charmed meson, and the corresponding partial widths are 38.4, 31.1, 25.4, 16.3 and 14.1 MeV, respectively. As for other $2F$ states,  including $2^3F_4$, $2F_3$, and $2F^\prime_3$ states, their decay behaviors are also investigated in the present work. The mass of the $2^3F_4$ charmed meson is calculated to be 3345 MeV in the modified GI model \cite{Song:2015fha}. The total and partial widths of the $2^3F_4$ charmed meson are also listed in Table \ref{Tab:decay}. The total width of $2^3F_4$ charmed meson is 155 MeV, which is about half of that of the $2^3F_2$ charmed meson. The main decay channels of the $2^3F_4$ charmed meson are $D_2^\ast(2460) \rho$, $D^* \rho$, $D^* a_2(1320)$ and $D^* f_2(1270)$, and the corresponding partial widths are 21.4, 18.9, 18.0, and 15.8 MeV, respectively.

\begin{table*}[htb]
\caption{A summary of the masses and decay behaviors of the $3P$ and $2F$ charmed mesons. 
The numbers in the brackets after the $D\pi $ channel are the corresponding branching ratios for different assignments. \label{Tab:Summary}}
\begin{tabular}{ccccc}
\toprule[1pt]\toprule[1pt]
&  Mass (MeV) & Width (MeV) & Main channels \\
\midrule[1pt]
Experiment & $3214 \pm 29 \pm 33 \pm 36$  & $186 \pm 38 \pm 34 \pm 63 $ & $D\pi$ \footnote{The $D\pi$ channel is the observed channel of the $D_2^\ast(3000)$ \cite{Aaij:2016fma}.} \\
$3^3P_2$  & 3234 & 102.4  & $D^\ast \pi$, $D \rho$, $D^\ast \rho$, $D_2^\ast(2460)\pi$, $D\pi(4.18\%)$  \\
$2^3F_2$  & 3364 & 302.2 & $D(1D_2\pi)$, $D_1(2420) \pi$, $D^\ast \rho$, $D\pi(1300)$, $D\pi(4.76\%)$, $D^\ast a_0(980)$ \\
\midrule[1pt]
$3^3P_0$  & 3219 &251.1   & $D\pi(1300)$, $D_1(2420) \pi$, $Db_1(1235)$, $D(1D_2) \pi$, $D \pi$, $D(2550) \pi$\\
$3P(1^+)$ & 3200      &144     &$Da_2(1320)$, $D^\ast \pi$ \\
$3P^\prime(1^+)$&3245 &185    &$D_2^\ast(2460) \pi$, $D(1^3D_3) \pi$             \\
$2F(3^+)$ & 3335       &165    &$D_2^\ast(2460)\rho$, $D^\ast \rho$, $D \rho$, $D_2^\ast(2460) \pi$ \\
$2F^\prime(3^+)$  &3377  &248 &$D(1^3D_3) \pi$, $D_2^*(2460) \pi$, $D^* \pi$                \\
$2^3F_4$ &3345 &155        &$D_2^\ast(2460) \rho$, $D^* \rho$, $D^* a_2(1320)$, $D^* f_2(1270)$ \\
\bottomrule[1pt]\bottomrule[1pt]
\end{tabular}
\end{table*}

The $F$-wave mixing states $2F(3^+)$ and $2F^\prime(3^+)$  are the mixtures of the $2^3F_3$ and $2^1F_3$ states, and the mixing scheme is
\begin{eqnarray}
\left(
\begin{array}{c}
  |2F(3^+)\rangle \\
  |2F^\prime(3^+) \rangle
\end{array}\right)
=
\left(
\begin{array}{cc}
 \cos(\theta_{2F}) & \sin (\theta_{2F}) \\
 -\sin(\theta_{2F}) &\cos (\theta_{2F}) \\
\end{array}
\right)
\left(
\begin{array}{c}
  |2^1F_3 \rangle \\
  |2^3F_3 \rangle
\end{array}
\right).
\end{eqnarray}
In the heavy quark limit, the mixing angle is $\theta_{2F}$= -arcsin(2/$\sqrt{7}$)= $-49.1^\circ$ \cite{Godfrey:1986wj, Matsuki:2010zy}. In the modified GI model, the masses of the $2^1F_3$ and $2^3F_3$ state are 3353 and 3359 MeV, respectively. The $\theta_{2F}$ dependence of the total and partial widths of $2F(3^+)$ and $2F^\prime(3^+)$ are shown in Figs. \ref{Fig:2F} and \ref{Fig:2Fprime}, respectively. The total width of $2F(3^+)$ charmed meson is 165 MeV with the mixing angle in the heavy quark limit. The dominant decay channels are $D_2^\ast(2460)\rho$, $D^\ast \rho$, $D \rho$, and $D_2^\ast(2460) \pi$, and the corresponding widths are 14.5, 14, 13.8 and 8.7 MeV, respectively. In addition, we find the partial width of the $2F(3^+) \to D^\ast \rho$ is weakly dependent on the mixing angle $\theta_{2F}$. The $D^\ast a_2(1320) $, $D^\ast b_1(1235)$, $D_1(2430)\rho$, and $D^\ast a_0(980)$ are also important decay channels of the $2F(3^+)$ charmed meson, the partial widths of which are all about 5 MeV. As for the $2F^\prime(3^+)$ charmed meson, the total width is 248 MeV. Our calculations indicate that the main decay channels of $2F^\prime (3^+)$ charmed meson are $D(1^3D_3) \pi$, $D_2^*(2460) \pi$, $D^* \pi$ and the corresponding widths are 39.8, 37.1 and 23.7 MeV, respectively. The partial widths of $D^* a_2(1320)$, $D^* b_1(1235)$, and $D^* \rho$ are all about 15 MeV, and $D^* f_2(1270)$ is about 10 MeV, which also has a considerable contribution to the total width.

\section{Discussion and summary \label{sec3}}

In the present work, we have studied the masses and decay behaviors of the $3P$ and $2F$ charmed mesons  and have compared our results of the $3^3P_2$ and $2^3F_2$ states with the observation of the $D_2^\ast(3000)$, which was reported very recently by the LHCb Collaboration \cite{Aaij:2016fma}. The modified GI model has predicted masses of the $3^3P_2$ and $2^3F_2$ states which are 3234 and 3364 MeV, respectively \cite{Song:2015nia}. On the other hand, the observed mass of the $D_2^\ast(3000)$ is $3214 \pm 29 \pm 33 \pm 36$ MeV,  which is quite consistent with that of the $3^3P_2$ charmed meson and about 150 MeV below the mass of the $2^3F_2$ charmed meson.

To further test the possibilities of the $D_2^\ast(3000)$ as a $3^3P_2 $ or $2^3F_2$ charmed meson, we have estimated the decay behaviors of the $3^3P_2 $ and $2^3F_2$ charmed mesons via the QPC model, in which the wave functions of the charmed and charmed strange mesons are evaluated by the modified GI model \cite{Song:2015fha, Song:2015nia}. The total width of the $3^3P_2 $ charmed meson is estimated to be 102.4 MeV, which is consistent with the lower limit of the width of the $D_2^\ast(3000)$. As for the $2^3F_2$ charmed meson, the total width is predicted to be 302.2 MeV, which is about 35 MeV above the upper limit of the width of the $D_2^\ast(3000)$. In addition, our calculated branching ratios for the $D\pi $ channel of the $3^3P_2$ and $2^3F_2$ charmed mesons are $4.18\%$ and 4.70$\%$ , respectively, which indicates the $D\pi$ channel is an important channel both for $3^3P_2$ and $2^3F_2$ charmed mesons. In Table \ref{Tab:Summary}, we summarize our calculations of the masses and decay behaviors of the $3^3P_2$ and $2^3F_2$ charmed mesons and compare them with the experimental measurement of the $D_2^\ast(3000)$. From the calculation in the present work, we find the most possible assignment of the $D_2^\ast(3000)$ is the $3^3P_2$ charmed meson, but the other possibility, such as the$2^3F_2$ charmed meson, may not be completely excluded. The predicted decay behaviors of the $3^3P_2$ and $2^3F_2$ charmed mesons could be tested by future experiments in LHCb and the forthcoming Belle II, which would help us reveal the nature of the $D_2^\ast(3000)$.

Besides the $3^3P_2 $ and $2^3F_2$ charmed mesons, the decay behaviors of the other six $3P$ and $2F$ charmed mesons have been also studied in the present work. The total width of the $3^3P_0$ charmed mesons is predicted to be 251.1 MeV, while the widths of the $3P(1^+)$ and $3P^\prime (1^+)$ charmed mesons are $144$ and 185 MeV, respectively, with the heavy quark limit mixing angle. The most dominant decay channels of the $3^3P_0$, $3P(1^+)$ and $3P^\prime(1+)$ are $D\pi(1300)$, $Da_2(1320)$ and $D_2^\ast(2460) \pi$, respectively. As for the $2F$ charmed mesons, the total widths of the $2F(3^+)$, $2F^\prime (3^+)$ and $2^3F_4$ are 165, 248, and 155 MeV, respectively. The estimates of the masses and decay behaviors of these unobserved states in the present work are also listed in Table. \ref{Tab:Summary}, which could provide some fundamental information for searching for these charmed mesons in future experiments.

\section*{Acknowledgments}

We would like to thank Bing Chen for useful discussion. This project is supported by the National Natural Science
Foundation of China under Grants No. 11222547, No. 11175073, No. 11375240, No. 11035006, and No. 11547301, and the Fundamental Research Funds for the Central Universities. X. L. is also supported by National Program for Support of Top-Notch Young Professionals.

{\it Note added:} During the preparation of this work, we noticed a new theoretical work related to $D_2^\ast(3000)$, in which the $D_2^\ast(3000)$ is assigned as a $1^3F_2$ state \cite{Wang:2016ewb}


\begin{thebibliography}{99}

\bibitem{Aaij:2016fma}
  R.~Aaij {\it et al.} [LHCb Collaboration],
  arXiv:1608.01289 [hep-ex].

\bibitem{delAmoSanchez:2010vq}
  P.~del Amo Sanchez {\it et al.}  [BaBar Collaboration],
  Phys.\ Rev.\ D {\bf 82} (2010) 111101
  [arXiv:1009.2076 [hep-ex]].

\bibitem{Aaij:2013sza}
  R.~Aaij {\it et al.}  [LHCb Collaboration],
  JHEP {\bf 1309} (2013) 145

\bibitem{Aubert:2009wg}
  B.~Aubert {\it et al.}  [BaBar Collaboration],
  Phys.\ Rev.\ D {\bf 79} (2009) 112004


\bibitem{Link:2003bd}
  J.~M.~Link {\it et al.}  [FOCUS Collaboration],
  Phys.\ Lett.\ B {\bf 586} (2004) 11


\bibitem{Abe:2003zm}
  K.~Abe {\it et al.}  [Belle Collaboration],
  Phys.\ Rev.\ D {\bf 69} (2004) 112002


\bibitem{Aubert:2006zb}
  B.~Aubert {\it et al.}  [BaBar Collaboration],
  Phys.\ Rev.\ D {\bf 74} (2006) 012001

\bibitem{Anjos:1988uf}
  J.~C.~Anjos {\it et al.}  [Tagged Photon Spectrometer Collaboration],
  Phys.\ Rev.\ Lett.\  {\bf 62} (1989) 1717.


\bibitem{Albrecht:1985as}
  H.~Albrecht {\it et al.}  [ARGUS Collaboration],
  Phys.\ Rev.\ Lett.\  {\bf 56} (1986) 549.

\bibitem{Godfrey:1985xj}
  S.~Godfrey and N.~Isgur,
  Phys.\ Rev.\ D {\bf 32} (1985) 189.


\bibitem{Goity:1998jr}
  J.~L.~Goity and W.~Roberts,
  Phys.\ Rev.\ D {\bf 60} (1999) 034001

\bibitem{Di Pierro:2001uu}
  M.~Di Pierro and E.~Eichten,
  Phys.\ Rev.\ D {\bf 64} (2001) 114004


\bibitem{Bardeen:2003kt}
  W.~A.~Bardeen, E.~J.~Eichten and C.~T.~Hill,
  Phys.\ Rev.\ D {\bf 68}, 054024 (2003)

\bibitem{Matsuki:1997da}
  T.~Matsuki and T.~Morii,
  Phys.\ Rev.\ D {\bf 56}, 5646 (1997)

\bibitem{Matsuki:2007zza}
  T.~Matsuki, T.~Morii and K.~Sudoh,
  Prog.\ Theor.\ Phys.\  {\bf 117}, 1077 (2007)

\bibitem{Matsuki:2006rz}
  T.~Matsuki, T.~Morii and K.~Sudoh,
  Eur.\ Phys.\ J.\ A {\bf 31}, 701 (2007)

\bibitem{Matsuki:2011xp}
  T.~Matsuki and K.~Seo,
  Phys.\ Rev.\ D {\bf 85}, 014036 (2012)

\bibitem{Ebert:1997nk}
  D.~Ebert, V.~O.~Galkin and R.~N.~Faustov,
  Phys.\ Rev.\ D {\bf 57}, 5663 (1998)
  [Erratum-ibid.\ D {\bf 59}, 019902 (1999)]

\bibitem{Song:2015nia}
  Q.~T.~Song, D.~Y.~Chen, X.~Liu and T.~Matsuki,
  Phys.\ Rev.\ D {\bf 91} (2015) 054031

\bibitem{Song:2015fha}
  Q.~T.~Song, D.~Y.~Chen, X.~Liu and T.~Matsuki,
  Phys.\ Rev.\ D {\bf 92}, no. 7, 074011 (2015)


\bibitem{Bali:2005fu}
  G.~S.~Bali, H.~Neff, T.~Dussel, T.Lippert and K.~Schilling,
  Phys.\ Rev.\ D {\bf 71} (2005) 114513


\bibitem{Armoni:2008jy}
  A.~Armoni,
  Phys.\ Rev.\ D {\bf 78} (2008) 065017

\bibitem{Bigazzi:2008gd}
  F.~Bigazzi, A.~L.~Cotrone, C.~Nunez and A.~Paredes,
  Phys.\ Rev.\ D {\bf 78} (2008) 114012


\bibitem{Namekawa:2011wt}
  Y.~Namekawa {\it et al.}  [PACS-CS Collaboration],
  Phys.\ Rev.\ D {\bf 84}, 074505 (2011)



\bibitem{Micu:1968mk}
  L.~Micu,
  Nucl.\ Phys.\ B {\bf 10}, 521 (1969).


\bibitem{LeYaouanc:1972vsx}
  A.~Le Yaouanc, L.~Oliver, O.~Pene and J.~C.~Raynal,
  Phys.\ Rev.\ D {\bf 8}, 2223 (1973).


\bibitem{LeYaouanc:1973ldf}
  A.~Le Yaouanc, L.~Oliver, O.~Pene and J.-C.~Raynal,
  Phys.\ Rev.\ D {\bf 9}, 1415 (1974).


\bibitem{LeYaouanc:1974cvx}
  A.~Le Yaouanc, L.~Oliver, O.~Pene and J.~C.~Raynal,
  Phys.\ Rev.\ D {\bf 11}, 1272  (1975).


\bibitem{LeYaouanc:1977fsz}
  A.~Le Yaouanc, L.~Oliver, O.~Pene and J.-C.~Raynal,
  Phys.\ Lett.\ B {\bf 71}, 397 (1977).

\bibitem{LeYaouanc:1977gm}
  A.~Le Yaouanc, L.~Oliver, O.~Pene and J.~C.~Raynal,
  Phys.\ Lett.\  B {\bf 72}, 57 (1977).



\bibitem{Hayne:1981zy}
  C.~Hayne and N.~Isgur,
  Phys.\ Rev.\ D {\bf 25}, 1944 (1982).



\bibitem{Jacob:1959at}
  M.~Jacob and G.~C.~Wick,
  Annals Phys.\  {\bf 7}, 404 (1959)
  [Annals Phys.\  {\bf 281}, 774 (2000)].










\bibitem{Agashe:2014kda}
  K.~A.~Olive {\it et al.} [Particle Data Group Collaboration],
  Chin.\ Phys.\ C {\bf 38}, 090001 (2014).


\bibitem{Yu:2011ta}
  J.~S.~Yu, Z.~F.~Sun, X.~Liu and Q.~Zhao,
  Phys.\ Rev.\ D {\bf 83} (2011) 114007

\bibitem{Godfrey:1986wj}
  S.~Godfrey and R.~Kokoski,
  Phys.\ Rev.\ D {\bf 43} (1991) 1679.




















\bibitem{Matsuki:2010zy}
  T.~Matsuki, T.~Morii and K.~Seo,
  Prog.\ Theor.\ Phys.\  {\bf 124} (2010) 285

\bibitem{Barnes:2002mu}
  T.~Barnes, N.~Black and P.~R.~Page,
  Phys.\ Rev.\ D {\bf 68} (2003) 054014



\bibitem{Wang:2016ewb}
  Z.~G.~Wang,
  arXiv:1608.02176 [hep-ph].



\end{thebibliography}
\end{document}